\documentclass[11pt,a4paper]{article}

\usepackage{amssymb}
\usepackage[dvips]{graphicx}
\usepackage{bm}

\begin{document}

\title{\bf One-loop divergences in non-Abelian
supersymmetric theories regularized by BRST-invariant version of
the higher derivative regularization}

\author{
S.S.Aleshin, A.E.Kazantsev, M.B.Skoptsov, K.V.Stepanyantz\\
{\small{\em Moscow State University}}, {\small{\em  Physical
Faculty, Department  of Theoretical Physics}}\\
{\small{\em 119991, Moscow, Russia}}}

\maketitle

\begin{abstract}
We consider a general non-Abelian renormalizable ${\cal N}=1$
supersymmetric gauge theory, regularized by higher covariant
derivatives without breaking the BRST invariance, and calculate
one-loop divergences for a general form of higher derivative regulator
and of the gauge fixing term. It is demonstrated that the momentum integrals
giving the one-loop $\beta$-function are integrals of double total
derivatives independently of a particular choice of the higher derivative term.
Evaluating them we reproduce the well-known result for the one-loop
$\beta$-function. Also we find that the three-point ghost vertices with a single
line of the quantum gauge superfield are not renormalized in the considered approximation.
\end{abstract}

\unitlength=1cm

keywords: higher covariant derivative regularization, supersymmetry, renormalization.

\section{Introduction}
\hspace*{\parindent}

The well-known non-renormalization theorem \cite{Grisaru:1979wc}
states that in ${\cal N}=1$ supersymmetric gauge theories the
superpotential does not receive divergent quantum corrections.
Moreover, there is one more interesting feature of quantum
corrections in these theories. Namely, the renormalization of the
coupling constant is related to the renormalization of the matter
superfields by the so-called exact NSVZ $\beta$-function
\cite{Novikov:1983uc,Jones:1983ip,Novikov:1985rd,Shifman:1986zi,Vainshtein:1986ja,Shifman:1985fi}.
For the general ${\cal N}=1$ supersymmetric Yang--Mills (SYM) theory
with matter it is written as

\begin{equation}
\beta(\alpha,\lambda) = - \frac{\alpha^2\Big(3 C_2 - T(R) +
C(R)_i{}^j (\gamma_\phi)_j{}^i(\alpha,\lambda)/r\Big)}{2\pi(1-
C_2\alpha/2\pi)},
\end{equation}

\noindent where $(\gamma_\phi)_j{}^i$ denotes the anomalous dimension of
the chiral matter superfields and the following notation is used:

\begin{eqnarray}
&& \mbox{tr}\,(T^A T^B) \equiv T(R)\,\delta^{AB};\qquad
(T^A)_i{}^k
(T^A)_k{}^j \equiv C(R)_i{}^j;\qquad\ \nonumber\\
&& f^{ACD} f^{BCD} \equiv C_2 \delta^{AB};\qquad\quad r \equiv
\delta_{AA}.\qquad
\end{eqnarray}

\noindent (It is assumed that the generators of the fundamental
representation $t^A$ are normalized by the condition
$\mbox{tr}(t^A t^B) = \delta^{AB}/2$.)

In the early papers the NSVZ $\beta$-function was obtained from
general arguments, such as the structure of instanton contributions
to the effective action \cite{Novikov:1983uc,Shifman:1999mv},
anomalies \cite{Jones:1983ip,Shifman:1986zi,ArkaniHamed:1997mj},
non-renormalization of the topological term \cite{Kraus:2002nu}.
However, it was also necessary to construct the subtraction scheme
in which the NSVZ $\beta$-function is valid. The explicit loop
calculations
\cite{Avdeev:1981ew,Jack:1996vg,Jack:1996cn,Harlander:2006xq,Jack:2007ni}
(see \cite{Mihaila:2013wma} for a review) made with the dimensional
reduction \cite{Siegel:1979wq} in the $\overline{\mbox{DR}}$-scheme
agree with the NSVZ $\beta$-function only after a special finite
renormalization which should be constructed in each order
\cite{Jack:1996vg,Jack:1998uj}. Up to now, there is no general
prescription how to do it in an arbitrary order. However, a
possibility of making this finite renormalization is nontrivial
\cite{Jack:1996vg}, because from the general equation which
describes how the NSVZ expression is changed under a finite
renormalization \cite{Kutasov:2004xu,Kataev:2014gxa} one can derive
some scheme independent consequences of the NSVZ relation
\cite{Kataev:2014gxa,Kataev:2013csa}.

At least in the Abelian case, the NSVZ scheme can be naturally
constructed if the supersymmetric theories are regularized by higher
covariant derivatives \cite{Slavnov:1971aw,Slavnov:1972sq}. This
regularization is mathematically consistent unlike the dimensional
reduction \cite{Siegel:1980qs}. (Removing the inconsistencies of the
dimensional reduction leads to the loss of the explicit
supersymmetry \cite{Avdeev:1981vf}, which can be in this case broken
by higher order quantum corrections
\cite{Avdeev:1982xy,Avdeev:1982np,Velizhanin:2008rw}.) The higher
covariant derivative regularization can be formulated in the
explicitly ${\cal N}=1$ supersymmetric way
\cite{Krivoshchekov:1978xg,West:1985jx}, so that it does not break
supersymmetry. It can be also used for regularization of ${\cal
N}=2$ supersymmetric theories
\cite{Krivoshchekov:1985pq,Buchbinder:2014wra,Buchbinder:2015eva}.

With the higher covariant derivative regularization the NSVZ
relation was derived for the Abelian supersymmetric theories in all
orders for the renormalization group (RG) functions defined in terms
of the bare coupling constant
\cite{Stepanyantz:2011jy,Stepanyantz:2014ima} (which are
scheme-independent for a fixed regularization
\cite{Kataev:2013eta}). The RG functions defined in the standard way
in terms of the renormalized coupling constant
\cite{Bogolyubov:1980nc} satisfy the NSVZ relation only in the NSVZ
scheme which in this case can be constructed in all orders by
imposing simple boundary conditions on the renormalization constants
\cite{Kataev:2013eta,Kataev:2014gxa}. Thus, the NSVZ scheme can be
easily constructed with the Slavnov higher derivative
regularization. The main feature of quantum corrections, which
allows to do this, is the factorization of integrals for the
$\beta$-function (defined in terms of the bare coupling constant)
into integrals of (double) total derivatives in the momentum space
in the limit of the vanishing external momentum
\cite{Soloshenko:2003nc,Smilga:2004zr}. In the Abelian case this has
been proved in all orders
\cite{Stepanyantz:2011jy,Stepanyantz:2014ima} and confirmed by
explicit calculations in the three-loop approximation
\cite{Kazantsev:2014yna}. Using a similar method it has been proved
that the integrals for the Adler $D$-function \cite{Adler:1974gd}
(defined in terms of the bare coupling constant) in ${\cal N}=1$
supersymmetric QCD are also integrals of double total derivatives in
all orders. This feature allows to relate this function to the
anomalous dimension of the matter superfields exactly in all orders
\cite{Shifman:2014cya,Shifman:2015doa}. This new relation is similar
to the NSVZ $\beta$-function and has a similar origin.

In the non-Abelian case the calculations of the $\beta$-function
with the higher covariant derivative regularization were made only
in the two-loop order \cite{Pimenov:2009hv,Stepanyantz_MIAN}, where
it was demonstrated that all momentum integrals giving the
$\beta$-function are integrals of total derivatives. Subsequently,
the results of the papers \cite{Pimenov:2009hv,Stepanyantz_MIAN}
were written in the form of integrals of double total derivatives
\cite{Stepanyantz:2011bz,Stepanyantz:2012zz,Stepanyantz:2012us}.
However, the versions of the higher covariant derivative
regularization which were used for making explicit calculations for
the non-Abelian ${\cal N}=1$ supersymmetric theories break the BRST
invariance \cite{Becchi:1974md,Tyutin:1975qk} (while the background gauge invariance is not broken).
Then the calculations are much simpler in comparison with the
version of the higher derivative regularization which does not break
the BRST invariance. Certainly, using of non-invariant
regularizations is possible (see, e.g.,
\cite{Slavnov:2001pu,Slavnov:2002ir,Slavnov:2002kg,Slavnov:2003cx}),
if they are supplemented by a subtraction scheme which restores the
Slavnov--Taylor identities \cite{Taylor:1971ff,Slavnov:1972fg}.
However, it is much more convenient to make calculations with the
invariant regularization. Moreover, the invariant regularization may
be useful for the general derivation of the NSVZ relation in the
non-Abelian case. That is why in the present paper we consider a
more complicated version of the higher derivative regularization
which does not break the BRST invariance and a very general forms of
the higher derivative term and of the gauge fixing term. In this case
the calculations are much more complicated. That is why here we make
them only in the one-loop approximation. Certainly, in the one-loop
approximation the higher derivative regularization always gives the
result which is in agreement with other regularizations
\cite{Pronin:1997eb}. Nevertheless, the one-loop calculations can
be used for demonstrating the factorization of integrals which give
the $\beta$-function into integrals of double total derivatives.
Moreover, they allow to verify the method of calculations and fix
some potential problems. For example, the first calculation of the
quantum corrections made with the higher covariant derivative
regularization for the (non-supersymmetric) Yang--Mills theory
\cite{Martin:1994cg} gave the correct result for the one-loop
$\beta$-function \cite{Gross:1973id,Politzer:1973fx} only after
corrections made in \cite{Asorey:1995tq,Bakeyev:1996is}. (One-loop
quantum  corrections in non-supersymmetric electrodynamics with the
higher derivative term were also recently investigated in
\cite{Turcati:2016aog}.) There are also other subtleties in calculating
quantum corrections in supersymmetric theories, see, e.g.,
\cite{ArkaniHamed:1997mj,Fargnoli:2010mf,Cherchiglia:2015vaa}.
One more important reason for making the
one-loop calculation is that for deriving the NSVZ relation by the
direct summation of supergraphs in all orders (such as in Refs.
\cite{Stepanyantz:2011jy,Stepanyantz:2014ima}) this approximation
should be considered separately and the BRST invariant
regularization is highly desirable.

This paper is organized as follows. In Sect. \ref{Section_SYM} we
regularize the ${\cal N}=1$ SYM theory by higher covariant
derivatives without breaking the BRST invariance and construct the
generating functional for the regularized theory. In Sect.
\ref{Section_One_Loop} we calculate the one-loop divergences for
various Green functions. In particular, we demonstrate that all
integrals giving the one-loop $\beta$-function are integrals of
double total derivatives in the momentum space independently of the
form of the higher derivative term, and ghost vertices with a single
gauge line are finite.

\section{The BRST-invariant higher covariant derivative
regularization for ${\cal N}=1$ supersymmetric gauge
theories}\label{Section_SYM}

\subsection{Action of the considered theory}
\hspace*{\parindent}\label{Subsection_SYM_Action}

In this paper we consider the general renormalizable ${\cal N}=1$
SYM theory. In the massless limit this theory is described by the
action

\begin{eqnarray}\label{Action_Classical}
&& S = \frac{1}{2 e_0^2}\,\mbox{Re}\,\mbox{tr}\int d^4x\,
d^2\theta\,W^a W_a + \frac{1}{4} \int d^4x\, d^4\theta\,\phi^{*i}
(e^{2V})_i{}^j \phi_j\nonumber\\
&&\qquad\qquad\qquad\qquad\qquad\qquad\qquad\qquad\qquad  +
\Big(\frac{1}{6} \int d^4x\,d^2\theta\,\lambda_0^{ijk} \phi_i
\phi_j \phi_k + \mbox{c.c.}\Big),\qquad
\end{eqnarray}

\noindent which is written in terms of ${\cal N}=1$ superfields
\cite{West:1990tg,Buchbinder:1998qv}. Here $e_0$ and
$\lambda^{ijk}_0$ denote the bare coupling constant and the Yukawa
couplings, respectively. The gauge superfield

\begin{equation}
V = e_0 V^A T^A
\end{equation}

\noindent is hermitian, so that its components, $V^A$, are real
superfields. $\phi_i$ are chiral matter superfields which lie in a
certain representation $R$ of the gauge group $G$. In general, this
representation can be reducible. The gauge superfield strength

\begin{equation}
W_a \equiv \frac{1}{8} \bar D^2 (e^{-2V} D_a e^{2V}) = e_0 W_a^A t^A
\end{equation}

\noindent is also a chiral superfield. In order to obtain a gauge
invariant theory, the Yukawa couplings should satisfy the condition

\begin{equation}
\lambda_0^{ijm} (T^A)_m{}^{k} + \lambda_0^{imk} (T^A)_m{}^{j} +
\lambda_0^{mjk} (T^A)_m{}^{i} = 0.
\end{equation}

\noindent In this case the theory (\ref{Action_Classical}) is
invariant under the transformations

\begin{equation}
\phi \to e^{A}\phi;\qquad e^{2V} \to e^{-A^+} e^{2V} e^{-A},
\end{equation}

\noindent where the parameter $A$ is an arbitrary chiral
superfield. Under these transformations the gauge superfield
strength is changed as

\begin{equation}
W_a \to e^{A} W_a e^{-A}.
\end{equation}

\noindent Note that in the last two equations we use the matrix
notation. Explicitly writing the indexes we obtain, e.g.,

\begin{equation}
\phi_i \to (e^{A})_i{}^j \phi_j,\qquad \mbox{etc}.
\end{equation}

\noindent It is also convenient to introduce the superfield
$\Omega$ which, by definition, satisfies the equation

\begin{equation}
e^{2V} \equiv e^{\Omega^+} e^{\Omega}.
\end{equation}

\subsection{The background field method}
\hspace*{\parindent}\label{Subsection_SYM_Background}

A convenient tool for calculating quantum corrections is the
background field method
\cite{DeWitt:1965jb,Abbott:1980hw,Abbott:1981ke}, because it
allows to obtain the explicitly gauge invariant effective action.
In the supersymmetric case it is introduced by the substitution

\begin{equation}\label{Background_Splitting}
e^{\Omega} \to e^{\Omega} e^{\bm{\Omega}},\qquad\mbox{so
that}\qquad e^{2V} \to e^{\bm{\Omega}^+} e^{2V} e^{\bm{\Omega}}.
\end{equation}

\noindent Then the background gauge superfield $\bm{V}$ is defined
by the equation

\begin{equation}
e^{2\bm{V}} = e^{\bm{\Omega}^+} e^{\bm{\Omega}}.
\end{equation}

\noindent The theory which is obtained after the substitution
(\ref{Background_Splitting}) is evidently invariant under the
background gauge transformations

\begin{equation}\label{Background_Invariance}
e^{\bm{\Omega}} \to e^{iK} e^{\bm{\Omega}} e^{-A};\qquad e^{\Omega}
\to e^{\Omega} e^{-iK};\qquad V\to e^{iK} V e^{-iK};\qquad \phi \to
e^A \phi,
\end{equation}

\noindent where $K$ is an arbitrary hermitian superfield and $A$
is a chiral superfield which lies in the Lie algebra of the gauge
group.

Also the considered theory is invariant under the quantum gauge
transformations

\begin{equation}\label{Quantum_Invariance}
e^{2V} \to e^{-{\cal A}^+} e^{2V} e^{-{\cal A}};\qquad
e^{\bm{\Omega}} \to e^{\bm{\Omega}};\qquad e^{\bm{\Omega^+}} \to
e^{\bm{\Omega^+}};\qquad \phi\to e^{-\bm{\Omega}} e^{\cal A}
e^{\bm{\Omega}}\phi.
\end{equation}

\noindent The parameter ${\cal A}$ of the quantum gauge
transformations is a background chiral superfield which, by
definition, satisfies the condition

\begin{equation}
\bm{\bar\nabla}_{\dot a} {\cal A} = 0,
\end{equation}

\noindent where the gauge and supersymmetric background covariant
derivatives are defined by

\begin{equation}
\bm{\nabla}_a = e^{-\bm{\Omega}^+} D_a e^{\bm{\Omega}^+}; \qquad
\bm{\bar\nabla}_{\dot a} = e^{\bm{\Omega}} \bar D_{\dot a}
e^{-\bm{\Omega}}.
\end{equation}

\noindent Acting on a superfield $S$ which transforms as $S \to
e^{iK} S$ under the background gauge symmetry (\ref{Background_Invariance})
they will have the same transformation law, e.g.,
$\bm{\nabla}_a S \to e^{iK} \bm{\nabla}_a S$.

After the substitution (\ref{Background_Splitting}) the superfield
strength $W_a$ will have the form

\begin{equation}\label{Total_W}
W_a = \frac{1}{8} e^{-\bm{\Omega}} \bm{\bar\nabla}^2 (e^{-2V}
\bm{\nabla}_a e^{2V}) e^{\bm{\Omega}} + \bm{W}_a,
\end{equation}

\noindent where we introduce the notation

\begin{equation}
\bm{W}_a = \frac{1}{8} \bar D^2 (e^{-2\bm{V}} D_a e^{2\bm{V}}).
\end{equation}

It is convenient to fix a gauge and introduce a regularization in
such a way that the invariance (\ref{Background_Invariance}) remains
unbroken. Then the effective action will be also invariant under the
transformations (\ref{Background_Invariance}), which is very
convenient for calculating RG functions.

\subsection{The higher covariant derivative regularization}
\hspace*{\parindent}\label{Subsection_SYM_HD}

The main idea of the higher covariant derivative regularization is
adding a term with higher degrees of the covariant derivatives
(which we will denote by $S_\Lambda$) to the classical action.
Certainly, such a term is not uniquely defined. There are a lot of
options for choosing it. Here, for definiteness, we will use the
following expression:

\begin{eqnarray}\label{Action_HD}
&& S_{\Lambda} = \frac{1}{2e_0^2}\,\mbox{Re}\,\mbox{tr}\int d^4x\,
d^2\theta\, e^{\Omega} e^{\bm{\Omega}} W^a e^{-\bm{\Omega}}
e^{-\Omega} \Big[R\Big(-\frac{\bar\nabla^2
\nabla^2}{16\Lambda^2}\Big) -1\Big]_{Adj} e^{\Omega} e^{\bm{\Omega}}
W_a
e^{-\bm{\Omega}} e^{-\Omega}\qquad\nonumber\\
&& + \frac{1}{4} \int d^4x\, d^4\theta\,\phi^+ e^{\bm{\Omega}^+}
e^{\Omega^+} \Big[F\Big(-\frac{\bar\nabla^2
\nabla^2}{16\Lambda^2}\Big)-1\Big] e^{\Omega} e^{\bm{\Omega}}
\phi,
\end{eqnarray}

\noindent where the gauge and supersymmetric covariant derivatives
here are given by

\begin{equation}
\nabla_a = e^{-\Omega^+} e^{-\bm{\Omega}^+} D_a e^{\bm{\Omega}^+}
e^{\Omega^+}; \qquad \bar\nabla_{\dot a} = e^{\Omega}
e^{\bm{\Omega}} \bar D_{\dot a} e^{-\bm{\Omega}} e^{-\Omega}.
\end{equation}

\noindent The subscript $Adj$ points out that they act on
superfields in the adjoint representation. In particular, this
implies that

\begin{equation}
\Big(f_0 + f_1 V + f_2 V^2 +\ldots\Big)_{Adj} X \equiv f_0 X + f_1
[V,X] + f_2 [V, [V, X]] + \ldots
\end{equation}

\noindent The functions $R(x)$ and $F(x)$ satisfy the conditions
$R(0)=1$, $F(0)=1$ and have polynomial growth in the limit $x\to
\infty$.

One can verify that the expression (\ref{Action_HD}) is invariant both
under the background gauge transformations (\ref{Background_Invariance})
and under the quantum gauge transformations (\ref{Quantum_Invariance}).
Taking into account that

\begin{equation}
- \frac{\bar D^2 D^2}{16\Lambda^2} \phi =
\frac{\partial^2}{\Lambda^2} \phi
\end{equation}

\noindent for an arbitrary chiral superfield $\phi$, we see that
in the lowest order in the (background and quantum) gauge
superfields the regulators give $R(\partial^2/\Lambda^2)-1$ and
$F(\partial^2/\Lambda^2)-1$. That is why Eq. (\ref{Action_HD}) is
really a supersymmetric higher derivative term. After adding
$S_\Lambda$ to the classical action $S$ we obtain the regularized
action

\begin{equation}
S_{\mbox{\scriptsize reg}} = S + S_\Lambda.
\end{equation}

\noindent It is convenient to choose the gauge fixing term in the
form

\begin{equation}
S_{\mbox{\scriptsize gf}} = -\frac{1}{16 \xi_0 e_0^2}\mbox{tr} \int
d^4x\,d^4\theta\,\bm{\nabla}^2 V K\Big(-\frac{\bm{\bar\nabla}^2
\bm{\nabla}^2}{16\Lambda^2}\Big)_{Adj}\bm{\bar \nabla}^2 V,
\end{equation}

\noindent because it does not break the background gauge
invariance (\ref{Background_Invariance}). Here the function
$K(x)$ by construction satisfies the conditions $K(0)=1$, $K(\infty)=\infty$, and $\xi_0$
is a constant. In this case the standard gauge fixing procedure
leads to the following actions for the Faddeev--Popov and
Nielsen--Kallosh ghosts:

\begin{eqnarray}
&& S_{\mbox{\scriptsize FP}} = \frac{1}{e_0^2} \mbox{tr} \int
d^4x\,d^4\theta\, \left(e^{\bm{\Omega}}\bar c e^{-\bm{\Omega}} +
e^{-\bm{\Omega}^+}\bar c^+ e^{\bm{\Omega}^+}\right)\nonumber\\
&&\qquad\qquad\qquad\quad \times \Big\{
\Big(\frac{V}{1-e^{2V}}\Big)_{Adj} \Big(e^{-\bm{\Omega}^+} c^+
e^{\bm{\Omega}^+}\Big)  + \Big(\frac{V}{1-e^{-2V}}\Big)_{Adj}
\Big(e^{\bm{\Omega}} c
e^{-\bm{\Omega}}\Big)\Big\};\qquad\\
&& S_{\mbox{\scriptsize NK}} = \frac{1}{2e_0^2}\mbox{tr}\int
d^4x\,d^4\theta\,b^+ \Big[e^{\bm{\Omega}^+}
K\Big(-\frac{\bm{\bar\nabla}^2 \bm{\nabla}^2}{16\Lambda^2}\Big)
e^{\bm{\Omega}}\Big]_{Adj} b.
\end{eqnarray}

\noindent
The ghost superfields $c = e_0 c^A t^A$, $\bar c = e_0 \bar c^A t^A$, $b = e_0 b^A t^A$ are
anticommuting and chiral. As usual, the Nielsen--Kallosh ghosts interact only with the background
gauge superfield $\bm{V}$ and, therefore, non-trivially contribute
only in the one-loop approximation.

Introducing an auxiliary (commuting) chiral superfield $f$ it is
also convenient to present the gauge fixing term in the form

\begin{eqnarray}\label{Gauge_Fixing_With_F}
&& S_{\mbox{\scriptsize gf}} = \frac{1}{e_0^2} \mbox{tr}\int
d^4x\,d^4\theta\,\Big(16 \xi_0\,  f^+ \Big[e^{\bm{\Omega^+}}
K^{-1}\Big(-\frac{\bm{\bar\nabla}^2 \bm{\nabla}^2}{16\Lambda^2}\Big)
e^{\bm{\Omega}}\Big]_{Adj} f \nonumber\\
&& \qquad\qquad\qquad\qquad\qquad\qquad\qquad\qquad +
e^{\bm{\Omega}} f e^{-\bm{\Omega}} \bm{\nabla}^2 V +
e^{-\bm{\Omega^+}} f^+ e^{\bm{\Omega^+}} \bm{\bar \nabla}^2 V
\Big).\qquad
\end{eqnarray}

One can easily verify that the sum

\begin{equation}\label{Action_With_Ghosts}
S+S_\Lambda+S_{\mbox{\scriptsize gf}} +S_{\mbox{\scriptsize FP}}
+S_{\mbox{\scriptsize NK}},
\end{equation}

\noindent which is obtained after the gauge fixing procedure, is
invariant under the background gauge transformations
(\ref{Background_Invariance}), under which $f$ and ghost superfields should be transformed
as

\begin{equation}
f \to e^A f e^{-A} = (e^A)_{Adj} f;\qquad c \to (e^A)_{Adj} c;\qquad \bar c \to (e^A)_{Adj} \bar c;\qquad b \to (e^A)_{Adj} b.
\end{equation}

\noindent However, it is evident that the action (\ref{Action_With_Ghosts})
is not invariant under the quantum gauge transformation
(\ref{Quantum_Invariance}). Instead of this invariance the total
action becomes invariant under the BRST transformations

\begin{eqnarray}
&& \delta V = - \varepsilon  \Big\{
\Big(\frac{V}{1-e^{2V}}\Big)_{Adj} \left(e^{-\bm{\Omega}^+} c^+
e^{\bm{\Omega}^+}\right)  + \Big(\frac{V}{1-e^{-2V}}\Big)_{Adj}
\left(e^{\bm{\Omega}} c e^{-\bm{\Omega}}\right)\Big\};\qquad
\delta \phi = \varepsilon c \phi;\nonumber\\
&& \delta \bar c = \varepsilon \bar D^2 (e^{-2\bm{V}} f^+
e^{2\bm{V}});\qquad\, \delta \bar c^+ = \varepsilon D^2(e^{2\bm{V}}f
e^{-2\bm{V}});\qquad \delta c = \varepsilon c^2;\qquad \delta c^+ =
\varepsilon (c^+)^2;
\vphantom{\frac{\Lambda^2}{\Lambda^2}}\quad\nonumber\\
&&\qquad\qquad\quad \delta f = \delta f^+ = 0; \qquad\quad \delta b
= \delta b^+ = 0; \qquad\quad \delta \bm{\Omega} =
\delta\bm{\Omega}^+ = 0,\vphantom{\frac{\Lambda^2}{\Lambda^2}}
\end{eqnarray}

\noindent where $\varepsilon\ne \varepsilon(x)$ is an anticommuting real scalar
parameter. Two first of these equations are equivalent to the equations
(\ref{Quantum_Invariance}) in which the parameter ${\cal A}$ is given by the expression

\begin{equation}\label{A_For_BRST}
{\cal A} = \varepsilon e^{\bm{\Omega}} c e^{-\bm{\Omega}};\qquad {\cal A}^+ = -\varepsilon e^{-\bm{\Omega}^+} c^+ e^{\bm{\Omega}^+}.
\end{equation}

\noindent This allows us to verify the nilpotency of the BRST
transformations. Really, the equalities $\delta_1 \delta_2 c = 0$
and $\delta_1\delta_2 \bar c = 0$ are evident, and

\begin{equation}
\delta_1 \delta_2 e^{2V} = \delta_1\Big(\varepsilon_2
e^{-\bm{\Omega}^+} c^+ e^{\bm{\Omega}^+} e^{2V} - e^{2V}
\varepsilon_2 e^{\bm{\Omega}} c e^{-\bm{\Omega}}\Big) =0.
\end{equation}

\noindent As a consequence, we obtain that $\delta_1\delta_2 V =
0$ and see that the BRST transformations are nilpotent. Writing
the gauge fixing term in the form (\ref{Gauge_Fixing_With_F}) and
using this property, one can easily verify the BRST invariance of
the action (\ref{Action_With_Ghosts}).

By introducing the higher derivative term one regularizes
divergences beyond the one-loop approximation
\cite{Faddeev:1980be}. In order to get rid of the remaining
one-loop divergences, it is necessary to insert the Pauli--Villars determinants
into the generating functional \cite{Slavnov:1977zf}. Due to the
absence of quadratic divergences in supersymmetric theories it is
possible to use the following Pauli--Villars determinants:

\begin{eqnarray}
&& Z[\bm{V},\mbox{Sources}] = \int DV\, D\phi\, Db\, D\bar c\, Dc\,
\mbox{Det}(PV,M_{\Phi}) \mbox{Det}(PV,M_{\varphi})^{-1}
\qquad\nonumber\\
&&\qquad\qquad\qquad\qquad\qquad\qquad\quad \times \exp\Big(iS + iS_\Lambda +
iS_{\mbox{\scriptsize gf}} + i S_{\mbox{\scriptsize FP}} + i
S_{\mbox{\scriptsize NK}} + i S_{\mbox{\scriptsize
source}}\Big),\qquad
\end{eqnarray}

\noindent where

\begin{equation}
\mbox{Det}(PV,M_{\Phi}) = \int D \Phi \exp\left(i
S_\Phi\right);\qquad \mbox{Det}(PV,M_{\varphi})^{-1} = \int
D\varphi \exp\left(i S_\varphi\right).
\end{equation}

\noindent Here $\Phi_i$ is an anticommuting superfield in the same
representation as $\phi_i$, three commuting superfields $\varphi_f$
lie in the adjoint representation of the gauge group, and  the
actions for the Pauli--Villars superfields are given by

\begin{eqnarray}\label{Pauli--Villars_Actions}
&&\hspace*{-6mm} S_\Phi = \frac{1}{4} \int
d^4x\,d^4\theta\,\Phi^{*i} \Big[e^{\bm{\Omega}^+} e^{\Omega^+}
F\Big(-\frac{\bar\nabla^2 \nabla^2}{16\Lambda^2}\Big) e^{\Omega}
e^{\bm{\Omega}}\Big]_i^{\ \ j} \Phi_j + \Big(\frac{1}{4} \int
d^4x\,d^2\theta\, (M_\Phi)^{ij} \Phi_i \Phi_j
+\mbox{c.c.}\Big);\nonumber\\
&&\hspace*{-6mm} S_\varphi = \frac{1}{2e_0^2} \mbox{tr} \int
d^4x\,d^4\theta\, \Big(\varphi_1^+ \Big[e^{\bm{\Omega}^+}
e^{\Omega^+} R\Big(-\frac{\bar\nabla^2 \nabla^2}{16\Lambda^2}\Big)
e^{\Omega} e^{\bm{\Omega}}\Big]_{Adj}\varphi_1 + \varphi_2^+
\Big[e^{\bm{\Omega}^+} e^{2V}
e^{\bm{\Omega}}\Big]_{Adj}\varphi_2 \nonumber\\
&&\hspace*{-6mm} + \varphi_3^+ \Big[e^{\bm{\Omega}^+} e^{2V}
e^{\bm{\Omega}}\Big]_{Adj}\varphi_3\Big) +
\frac{1}{2e_0^2}\mbox{tr}\Big(\int d^4x\,d^2\theta\,M_\varphi
\left(\varphi_1^2 + \varphi_2^2 + \varphi_3^2\right)
+\mbox{c.c.}\Big).
\end{eqnarray}

\noindent
We will also assume that

\begin{equation}\label{M_Phi_Definition}
(M_\Phi)^{ji} (M_\Phi^*)_{kj} = M_\Phi^2\, \delta_k^i.
\end{equation}

\noindent
Below we will see that the chiral scalar superfields
$\varphi_f$ introduced in this way exactly cancel the one-loop
(sub)divergences introduced by loops of the gauge superfield and
ghosts.

The actions (\ref{Pauli--Villars_Actions}) are also BRST invariant,
because they are evidently invariant under the quantum gauge
transformations (\ref{Quantum_Invariance}) if

\begin{equation}
\Phi \to e^{-\bm{\Omega}} e^{{\cal A}} e^{\bm{\Omega}} \Phi;\qquad
\varphi_f \to e^{-\bm{\Omega}} e^{{\cal A}} e^{\bm{\Omega}}
\varphi_f e^{-\bm{\Omega}} e^{-{\cal A}} e^{\bm{\Omega}} =
\left(e^{-\bm{\Omega}} e^{{\cal A}} e^{\bm{\Omega}}\right)_{Adj}
\varphi_f.
\end{equation}

\noindent For ${\cal A}$ given by Eq. (\ref{A_For_BRST}) we obtain
the BRST invariance.

Thus, the final expression for the generating functional can be
written as

\begin{equation}
Z[\bm{V},\mbox{Sources}] = \int D\mu\,
\exp\Big(iS_{\mbox{\scriptsize total}} + i S_{\mbox{\scriptsize
source}}\Big),
\end{equation}

\noindent where $\int D\mu$ denotes the integration measure which
includes integration over all superfields of the theory, and the
total action

\begin{equation}
S_{\mbox{\scriptsize total}} = S + S_\Lambda + S_{\mbox{\scriptsize
gf}} + S_{\mbox{\scriptsize FP}} + S_{\mbox{\scriptsize NK}} +
S_\Phi + S_\varphi
\end{equation}

\noindent is invariant under the above described BRST
transformations. The generating functional for the connected Green
functions is given by

\begin{equation}
W[\bm{V},\mbox{Sources}] = -i \ln Z[\bm{V},\mbox{Sources}],
\end{equation}

\noindent and the effective action $\Gamma[\bm{V},\mbox{Fields}]$ is
defined in the standard way by using the Legendre transformation.

\subsection{Renormalization}
\hspace*{\parindent}\label{Subsection_SYM_Renormalization}

It is well known \cite{Slavnov:1974dg,Ferrara:1975ye,Piguet:1975md,Piguet:1981hh} that the considered
supersymmetric theory is renormalizable, so that the divergences can
be absorbed into redefinitions of the coupling constants and
(super)fields. Taking into account that the superpotential does not
receive the divergent quantum corrections according to the
non-renormalization theorem \cite{Grisaru:1979wc} we can make the
following renormalization:

\begin{eqnarray}\label{Renormalization_Constants_Definition}
&& \frac{1}{\alpha_0} = \frac{Z_\alpha}{\alpha};\qquad\
\frac{1}{\xi_0} = \frac{Z_\xi}{\xi};\qquad\ \bm{V} =
\bm{V}_R;\qquad\  V = Z_V Z_\alpha^{-1/2} V_R;\qquad\  \bar c c = Z_c Z_\alpha^{-1} \bar c_R c_R;\nonumber\\
&& \phi_i = (\sqrt{Z_\phi})_i{}^j (\phi_R)_j;\qquad\ \lambda^{ijk} =
\lambda_0^{mnp} (\sqrt{Z_\phi})_m{}^i (\sqrt{Z_\phi})_n{}^j
(\sqrt{Z_\phi})_p{}^k;\qquad b = \sqrt{Z_b} b_R,\qquad
\end{eqnarray}

\noindent where $\alpha$ and $\lambda^{ijk}$ are the renormalized
coupling constant and Yukawa couplings, respectively, and the
renormalized superfields are denoted by the subscript $R$. We also
take into account that the background gauge superfield is not
renormalized due to the unbroken background gauge symmetry
(\ref{Background_Invariance}). From the definitions
(\ref{Renormalization_Constants_Definition}) we see that $Z_c$
denotes the renormalization constant for the Faddeev--Popov ghosts;
$Z_V$ is the renormalization constant for the quantum gauge
superfield, and $Z_\alpha$ encodes the charge renormalization.

Because the Nielsen--Kallosh action is not renormalized, the
renormalization constants satisfy the relation $Z_\alpha Z_b=1$.
Similarly, taking into account that the two-point Green function of
the quantum gauge superfield is transversal due to the
Slavnov--Taylor identity \cite{Taylor:1971ff,Slavnov:1972fg}, we
obtain the relation $Z_\xi Z_V^2 = 1$.

The renormalization constants $Z_\alpha$ and $Z_\phi$ can be found
by calculating the two-point Green functions of the background gauge
superfield $\bm{V}$ and the matter superfields, respectively. Due to the
background gauge invariance (\ref{Background_Invariance}) the
corresponding part of the effective action can be written in the
form

\begin{eqnarray}
&& \Gamma^{(2)}_{\bm{V},\phi} = - \frac{1}{8\pi}\mbox{tr} \int
\frac{d^4p}{(2\pi)^4}\,d^4\theta\,\bm{V}(\theta,-p)\,\partial^2\Pi_{1/2}
\bm{V}(\theta,p)\,
d^{-1}(\alpha_0,\lambda_0,\Lambda/p)\qquad\nonumber\\
&&\qquad\qquad\qquad\qquad\qquad\qquad + \frac{1}{4} \int
\frac{d^4p}{(2\pi)^4}\, d^4\theta\, \phi^{*i}(\theta,-p)
\phi_j(\theta,p) (G_\phi)_i{}^j(\alpha_0,\lambda_0,\Lambda/p),\qquad
\end{eqnarray}

\noindent where $\partial^2\Pi_{1/2} = -D^a \bar D^2 D_a/8$ denotes
the supersymmetric transversal projection operator. The functions
$d^{-1}$ and $(G_\phi)_i{}^j$ are, in general, divergent in the limit
$\Lambda\to\infty$. The renormalized coupling constant
$\alpha(\alpha_0,\lambda_0,\Lambda/\mu)$ and the renormalization
constant $(Z_\phi)_i{}^j(\alpha_0,\lambda_0,\Lambda/\mu)$, where $\mu$ is a renormalization point, are
defined by requiring finiteness of the expressions

\begin{eqnarray}
&& d^{-1}\Big(\alpha_0(\alpha,\lambda,\Lambda/\mu),
\lambda_0(\alpha,\lambda,\Lambda/\mu),\Lambda/p\Big);\nonumber\\
&& (Z_\phi)_i{}^j\,
(G_\phi)_j{}^k\Big(\alpha_0(\alpha,\lambda,\Lambda/\mu),
\lambda_0(\alpha,\lambda,\Lambda/\mu),\Lambda/p\Big)
\end{eqnarray}

\noindent (considered as functions of $\alpha$, $\lambda$, $\mu/p$
and $\Lambda/p$) in the limit $\Lambda\to \infty$. Then the
renormalization constant $Z_\alpha$ is obtained from the equation

\begin{equation}
Z_\alpha = \frac{\alpha}{\alpha_0}.
\end{equation}

In order to find the remaining renormalization constants $Z_V$ and
$Z_c$ we consider the two-point Green functions of the quantum gauge
superfield and ghosts. The Slavnov--Taylor identity ensures that all
quantum corrections to the Green function of the quantum gauge
superfield are transversal, so that

\begin{eqnarray}
&& \Gamma^{(2)}_{V,c} - S_{\mbox{\scriptsize gf}}^{(2)} = -
\frac{1}{2e_0^2}\mbox{tr} \int
\frac{d^4p}{(2\pi)^4}\,d^4\theta\,V(\theta,-p)\,\partial^2\Pi_{1/2}
V(\theta,p)\,
G_V(\alpha_0,\lambda_0,\Lambda/p)\qquad\nonumber\\
&& + \frac{1}{2e_0^2} \mbox{tr} \int \frac{d^4p}{(2\pi)^4}\,
d^4\theta\,\Big(-\bar c(\theta,-p) c^+(\theta,p) + \bar
c^+(\theta,-p) c(\theta,p) \Big)
G_c(\alpha_0,\lambda_0,\Lambda/p).\qquad
\end{eqnarray}

\noindent Then the renormalization constants $Z_V$ and $Z_c$ can be
obtained by requiring finiteness of the functions

\begin{eqnarray}
&& Z_V^2 G_V\Big(\alpha_0(\alpha,\lambda,\Lambda/\mu),
\lambda_0(\alpha,\lambda,\Lambda/\mu),\Lambda/p\Big)\qquad
\mbox{and}\nonumber\\
&& Z_c G_c\Big(\alpha_0(\alpha,\lambda,\Lambda/\mu),
\lambda_0(\alpha,\lambda,\Lambda/\mu),\Lambda/p\Big)
\end{eqnarray}

\noindent in the limit $\Lambda\to \infty$, respectively.

\subsection{RG functions}
\hspace*{\parindent}\label{Subsection_SYM_RG_Functions}

In this paper we consider the RG functions defined in terms of the
bare coupling constant. In particular, the $\beta$-function is
defined in terms of the bare coupling constants according to the
prescription

\begin{equation}
\beta(\alpha_0,\lambda_0) =
\frac{d\alpha_0(\alpha,\lambda,\Lambda/\mu)}{d\ln\Lambda}
\Big|_{\alpha,\lambda=\mbox{\scriptsize const}},
\end{equation}

\noindent and can be related to the renormalization constant
$Z_\alpha$. Really, differentiating the first equation in
(\ref{Renormalization_Constants_Definition}) with respect to
$\ln\Lambda$ we obtain

\begin{equation}
\beta(\alpha_0,\lambda_0) = - \alpha_0 \frac{d\ln
Z_\alpha}{d\ln\Lambda}\Big|_{\alpha,\lambda=\mbox{\scriptsize
const}}.
\end{equation}

\noindent For calculating this $\beta$-function it is convenient to
consider the expression

\begin{equation}\label{We_Will_Calculate}
\frac{d}{d\ln \Lambda}\,
\Big(d^{-1}(\alpha_0,\lambda_0,\Lambda/p)-\alpha_0^{-1}\Big)\Big|_{p=0}
= - \frac{d\alpha_0^{-1}(\alpha,\lambda,\Lambda/\mu)}{d\ln\Lambda} =
\frac{\beta(\alpha_0,\lambda_0)}{\alpha_0^2},
\end{equation}

\noindent in which the derivative with respect to $\ln\Lambda$ is
calculated at fixed values of the renormalized coupling constant
$\alpha$ and renormalized Yukawa constants $\lambda^{ijk}$ in the limit of the vanishing external momentum $p$.

The anomalous dimensions are defined in terms of the bare coupling
constants by the equations

\begin{eqnarray}\label{Bare_Anomalous_Dimensions}
&&\hspace*{-3mm} (\gamma_\phi)_i{}^j(\alpha_0,\lambda_0) \equiv -\frac{d \ln
(Z_\phi)_i{}^j(\alpha,\lambda,\Lambda/\mu)}{d\ln\Lambda}\Big|_{\alpha,\lambda=\mbox{\scriptsize
const}} = \frac{d \ln
(G_\phi)_i{}^j(\alpha_0,\lambda_0,\Lambda/p)}{d\ln\Lambda}\Big|_{\alpha,\lambda=\mbox{\scriptsize
const};\ p=0};\nonumber\\
&&\hspace*{-3mm} \gamma_V(\alpha_0,\lambda_0) \equiv -\frac{d \ln
Z_V(\alpha,\lambda,\Lambda/\mu)}{d\ln\Lambda}\Big|_{\alpha,\lambda=\mbox{\scriptsize
const}} = \frac{1}{2}\cdot \frac{d \ln
G_V(\alpha_0,\lambda_0,\Lambda/p)}{d\ln\Lambda}\Big|_{\alpha,\lambda=\mbox{\scriptsize
const};\ p=0};\nonumber\\
&&\hspace*{-3mm} \gamma_c(\alpha_0,\lambda_0) \equiv -\frac{d \ln
Z_c(\alpha,\lambda,\Lambda/\mu)}{d\ln\Lambda}\Big|_{\alpha,\lambda=\mbox{\scriptsize
const}} = \frac{d \ln G_c(\alpha_0,\lambda_0,\Lambda/p)}{d\ln\Lambda}\Big|_{\alpha,\lambda=\mbox{\scriptsize
const};\ p=0}.\qquad
\end{eqnarray}

It is known \cite{Kataev:2013eta} that the RG functions defined in
terms of the bare couplings depend on the regularization, but do not
depend on the subtraction scheme for a fixed regularization.

\section{The RG functions in the one-loop approximation}
\label{Section_One_Loop}

\subsection{One-loop $\beta$-function}
\hspace*{\parindent}\label{Subsection_One-Loop_Beta}

\begin{figure}[h]

\begin{picture}(0,4)
\put(0.6,2.4){\includegraphics[scale=0.4]{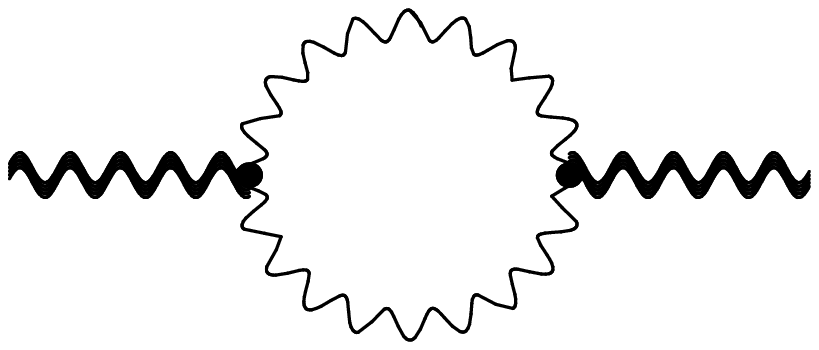}}
\put(1,0){\includegraphics[scale=0.4]{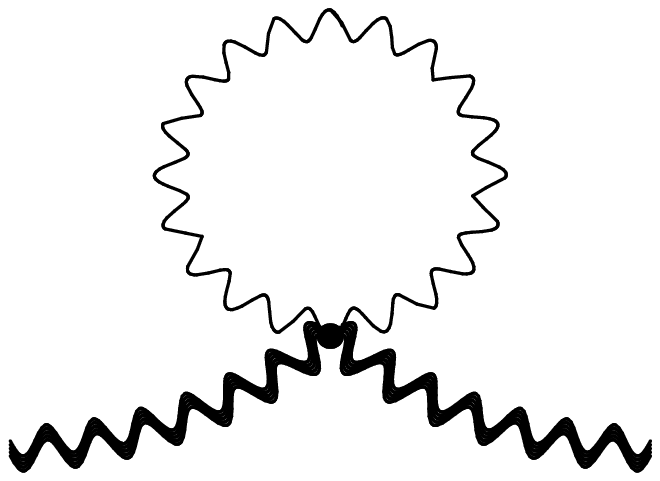}}
\put(4.4,2.43){\includegraphics[scale=0.4]{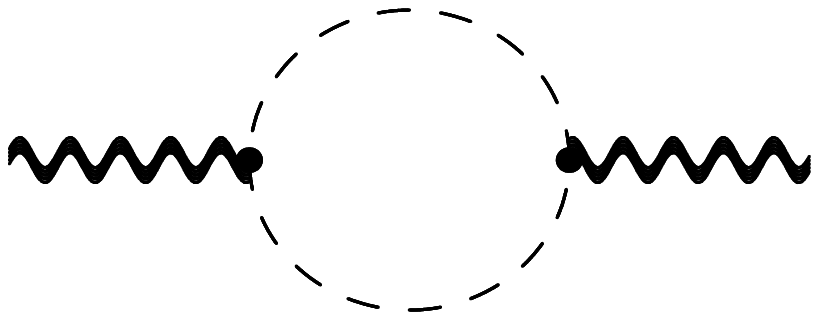}}
\put(4.7,-0.01){\includegraphics[scale=0.4]{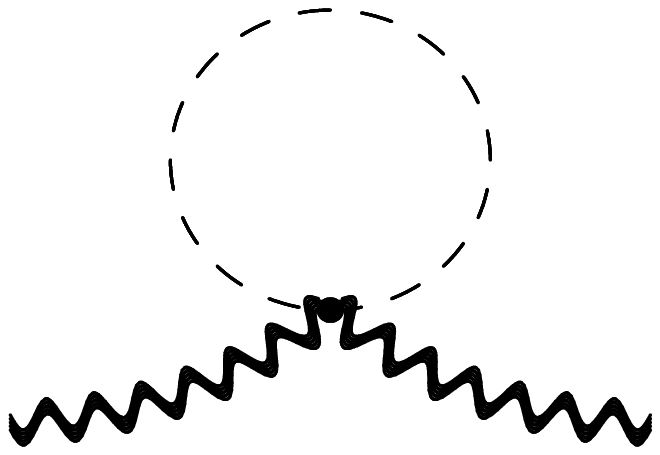}}
\put(8.2,2.4){\includegraphics[scale=0.4]{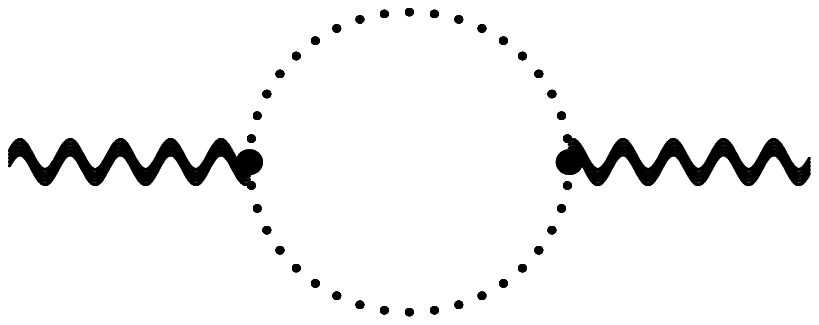}}
\put(8.6,-0.02){\includegraphics[scale=0.4]{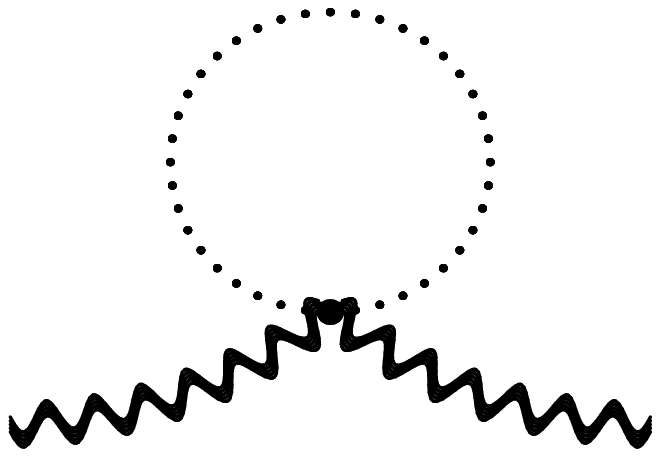}}
\put(12.1,2.4){\includegraphics[scale=0.4]{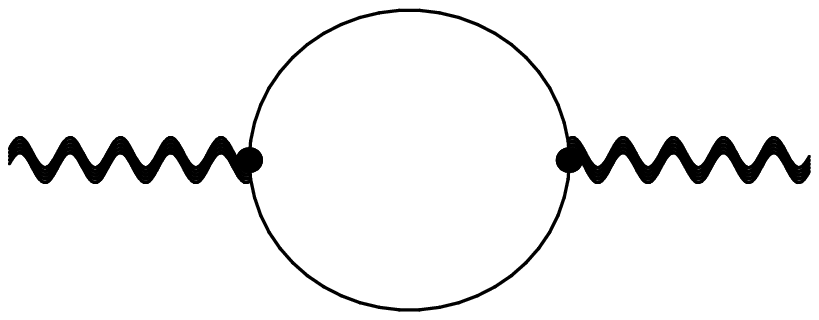}}
\put(12.45,-0.02){\includegraphics[scale=0.4]{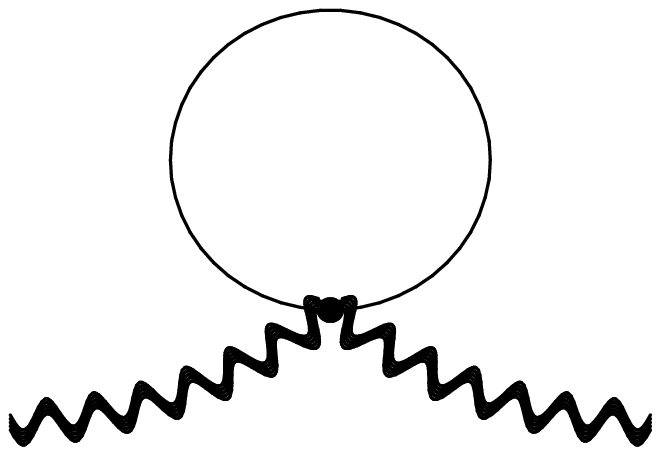}}
\end{picture}

\caption{One-loop diagrams which contribute to the two-point Green
function of the background superfield.}
\label{Figure_One_Loop_Diagrams}
\end{figure}

The two-point Green function of the background gauge superfield in
the one-loop approximation is contributed by the diagrams presented
in Fig. \ref{Figure_One_Loop_Diagrams}. In these diagrams external
lines correspond to the superfield $\bm{V}$. The wavy internal lines
denote propagators of the quantum gauge superfield $V$; the solid
lines denote propagators of the matter superfields $\phi_i$ and of
the Pauli--Villars superfields $\Phi_i$ and $\varphi_f$; the dashed
lines denote propagators of the Faddeev--Popov ghosts; the dotted
lines denote propagators of the Nielsen--Kallosh ghosts.

After calculating these diagrams we have obtained the following
result:

\begin{equation}\label{Beta_Original}
\frac{\beta(\alpha_0,\lambda_0)}{\alpha_0^2} = C_2 I_V + T(R) I_\phi
+ O(\alpha_0,\lambda_0^2).
\end{equation}

\noindent Here $I_V$ is the contribution of the quantum gauge
superfield, the (Faddeev--Popov and Nielsen--Kallosh) ghosts, and
the Pauli--Villars superfields $\varphi_f$. $I_\phi$ denotes the
contribution of the matter superfield $\phi$ and the corresponding
Pauli--Villars superfield $\Phi$. We have verified that both these
integrals are integrals of double total derivatives independently of
the concrete form of the functions $R$ and $F$ and have the
following form:

\begin{eqnarray}
&& I_V = \pi\int \frac{d^4q}{(2\pi)^4} \frac{d}{d\ln\Lambda}
\frac{\partial}{\partial q^\mu} \frac{\partial}{\partial q_\mu}
\Big[\frac{2}{q^2} \ln
\Big(\frac{R(q^2/\Lambda^2)}{K(q^2/\Lambda^2)}\Big)\nonumber\\
&& \qquad\qquad\qquad\qquad\qquad\qquad\quad - \frac{2}{q^2} \ln
\Big(1+ \frac{M_\varphi^2}{q^2}\Big) - \frac{1}{q^2}\ln\Big(
\frac{q^2 R^2(q^2/\Lambda^2)
+M_\varphi^2}{q^2 K^2(q^2/\Lambda^2)}\Big)\Big];\qquad\quad\\
&& I_{\phi} = \pi\int \frac{d^4q}{(2\pi)^4} \frac{d}{d\ln\Lambda}
\frac{\partial}{\partial q^\mu} \frac{\partial}{\partial q_\mu}
\Big[\frac{1}{q^2} \ln\Big(1 + \frac{M_\Phi^2}{q^2
F^2(q^2/\Lambda^2)}\Big)\Big],
\end{eqnarray}

\noindent
where $M_\Phi$ is defined by Eq. (\ref{M_Phi_Definition}). (The first
term in the integral $I_V$ is the contribution of diagrams with
the loop of the quantum gauge superfield $V$. The second term is
a sum of diagrams with the loop of the Faddeev--Popov ghosts and
the loop of the Pauli--Villars superfields $\varphi_2$ and $\varphi_3$.
The last term corresponds to diagrams with the loop of the
Nielsen--Kallosh ghosts and the Pauli--Villars superfield $\varphi_1$.)

We see that all these integrals are integrals of double total
derivatives in the momentum space. However, in general, they do not
vanish, because of singularities of the integrands. Really, let
$f(q^2/\Lambda^2)$ be a non-singular function with a rapid falloff
at infinity. Then we consider the integral of the double total
derivative

\begin{equation}
I\equiv \int \frac{d^4q}{(2\pi)^4} \frac{\partial}{\partial q^\mu}
\frac{\partial}{\partial q_\mu}\Big( \frac{1}{q^2}
f(q^2/\Lambda^2)\Big).
\end{equation}

\noindent This integral can be easily reduced to the integral of
the $\delta$-function singularity:

\begin{eqnarray}\label{Singular_Integral}
&& I = \int \frac{d^4q}{(2\pi)^4} \frac{\partial}{\partial q^\mu}
\Big(- \frac{2 q^\mu}{q^4} f(q^2/\Lambda^2) +
\frac{2q^\mu}{q^2\Lambda^2} f'(q^2/\Lambda^2)\Big)\nonumber\\
&& = \int \frac{d^4q}{(2\pi)^4} \frac{1}{q^2} \frac{d}{dq^2}
\Big(- 4 f(q^2/\Lambda^2) + \frac{4q^2}{\Lambda^2}
f'(q^2/\Lambda^2)\Big)\nonumber\\
&& = \frac{1}{4\pi^2}\Big( f(q^2/\Lambda^2) -
\frac{q^2}{\Lambda^2} f'(q^2/\Lambda^2)\Big)\Big|_{q=0} =
\frac{1}{4\pi^2} f(0) = 4\pi^2 \int \frac{d^4q}{(2\pi)^4}
\delta^4(q) f(q^2/\Lambda^2).\qquad
\end{eqnarray}

\noindent Using this result we calculate the integrals $I_V$ and
$I_\phi$. For example, let us consider the integral $I_\phi$. First,
we make the differentiation with respect to $\ln\Lambda$ taking into
account that $M_\Phi$ is proportional to $\Lambda$ and, then, use
Eq. (\ref{Singular_Integral}):

\begin{eqnarray}
&& I_{\phi} = \pi\int \frac{d^4q}{(2\pi)^4} \frac{\partial}{\partial
q^\mu} \frac{\partial}{\partial q_\mu}
\Bigg(\frac{2M_\Phi^2}{q^2\left(q^2 F^2(q^2/\Lambda^2) +
M_\Phi^2\right)} + \frac{4 M_\Phi^2 F'(q^2/\Lambda^2)}{\Lambda^2
F(q^2/\Lambda^2)\left(q^2 F^2(q^2/\Lambda^2) +
M_\Phi^2\right)}\Bigg)\quad\nonumber\\
&& = \int \frac{d^4q}{(2\pi)^4} \delta^4(q) \frac{8\pi^3
M_\Phi^2}{\left(q^2 F(q^2/\Lambda^2) + M_\Phi^2\right)} =
\frac{1}{2\pi}.
\end{eqnarray}

\noindent Similarly, we obtain

\begin{equation}
I_V = - \pi\int \frac{d^4q}{(2\pi)^4} \frac{d}{d\ln\Lambda}
\frac{\partial}{\partial q^\mu} \frac{\partial}{\partial q_\mu}
\Big[\frac{2}{q^2} \ln \Big(1+ \frac{M_\varphi^2}{q^2}\Big) +
\frac{1}{q^2}\ln\Big(1+ \frac{M_\varphi^2}{q^2 R^2} \Big)\Big] =
-\frac{3}{2\pi}.
\end{equation}

\noindent Therefore, in the one-loop approximation the
$\beta$-function (defined in terms of the bare coupling constant) is

\begin{equation}\label{Beta}
\beta = -\frac{\alpha_0^2}{2\pi}\Big(3C_2 - T(R) +
O(\alpha_0,\lambda_0^2)\Big).
\end{equation}

\noindent Thus, we reobtain the standard expression for the one-loop
$\beta$-function, which was first found in \cite{Ferrara:1974pu}.

In the end of this section we note that Eq. (\ref{Beta_Original})
can be also rewritten in the form

\begin{eqnarray}\label{Gamma_Alpha}
&& \frac{d\ln Z_\alpha}{d\ln\Lambda} = \pi \alpha_0 \int
\frac{d^4q}{(2\pi)^4} \frac{d}{d\ln\Lambda} \frac{\partial}{\partial
q^\mu} \frac{\partial}{\partial q_\mu} \Big[ \frac{C_2}{q^2}\ln
\Big(1+\frac{M_\varphi^2}{q^2 R^2}\Big)\nonumber\\
&&\qquad\qquad\qquad\qquad\qquad\quad +\frac{2C_2}{q^2}
\ln\Big(1+\frac{M_\varphi^2}{q^2}\Big) -
\frac{T(R)}{q^2}\ln\Big(1+\frac{M_\Phi^2}{q^2 F^2}\Big)\Big] +
O(\alpha_0^2,\alpha_0\lambda_0^2),\qquad
\end{eqnarray}

\noindent which will be useful below.

\subsection{One-loop anomalous dimension of the matter superfields}
\hspace*{\parindent}\label{Subsection_One-Loop_Gamma}

\begin{figure}[h]
\begin{picture}(0,2)
\put(2,0.2){\includegraphics[scale=0.45]{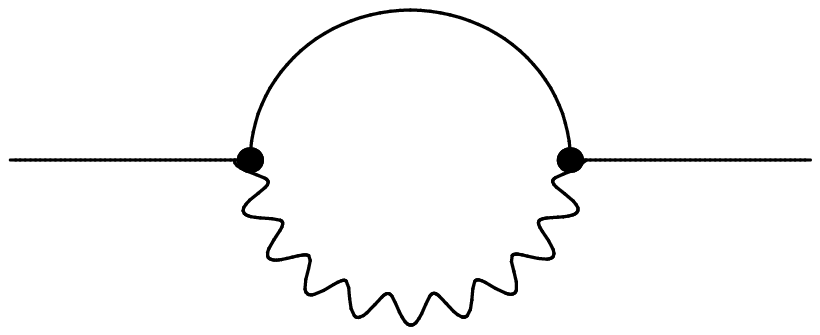}}
\put(6,-0.3){\includegraphics[scale=0.43]{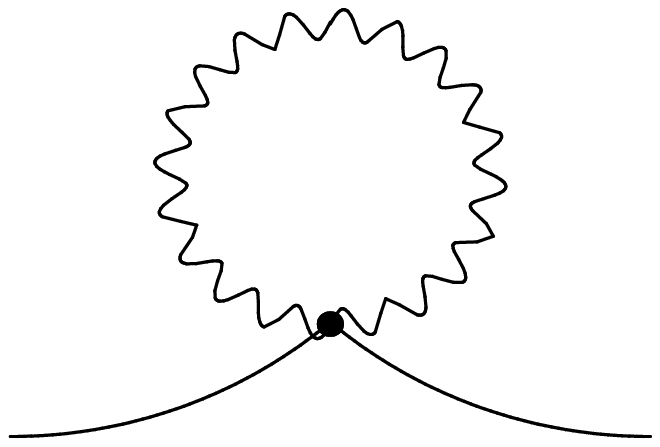}}
\put(10.1,0.2){\includegraphics[scale=0.45]{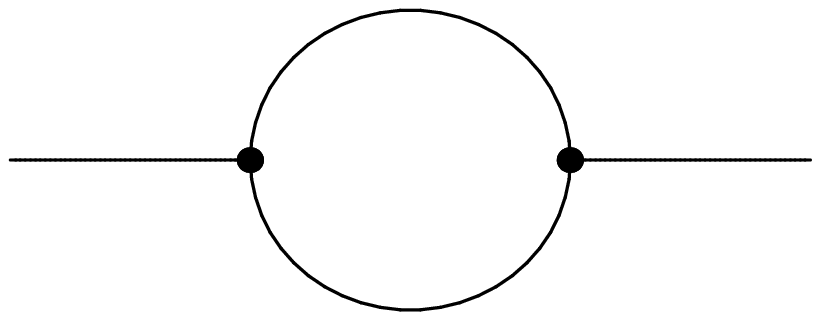}}
\end{picture}
\caption{Diagrams contributing to the one-loop anomalous dimension
of the matter superfield.}\label{Figure_Anomalous_Dimension}
\end{figure}

The one-loop anomalous dimension of the superfields $\phi_i$ for the considered theory is
determined by the diagrams presented in Fig.
\ref{Figure_Anomalous_Dimension}. They give the following result
for the anomalous dimension defined in terms of the bare coupling
constant:

\begin{eqnarray}\label{Anomalous_Dimension_Original}
&& (\gamma_\phi)_i{}^j(\alpha_0,\lambda_0) = \int \frac{d^4k}{(2\pi)^4}
\frac{d}{d\ln\Lambda} \Big(- C(R)_i{}^j \frac{2e^2}{k^4
R(k^2/\Lambda^2)} + \lambda^*_{imn} \lambda^{jmn} \frac{2}{k^4
F^2(k^2/\Lambda^2)} \Big)\qquad \nonumber\\
&& + O(\alpha^2,\alpha\lambda^2,\lambda^4).\vphantom{\frac{1}{2}}
\end{eqnarray}

\noindent Due to the derivative with respect to $\ln\Lambda$
(which should be taken at fixed values of the renormalized
coupling and Yukawa constants $e$ and $\lambda^{ijk}$,
respectively) this integral is well defined. Taking into account
that

\begin{equation}\label{Derivative}
\frac{d}{d\ln\Lambda} f(k^2/\Lambda^2) = - \frac{d}{d\ln k}
f(k^2/\Lambda^2) = - 2 k^2 \frac{d}{dk^2} f(k^2/\Lambda^2)
\end{equation}

\noindent and that for an arbitrary function $f$ with a sufficiently
rapid falloff at infinity

\begin{equation}
\int \frac{d^4k}{(2\pi)^4} \frac{1}{k^2} \frac{d f}{dk^2} =
-\frac{1}{16\pi^2} f(0),
\end{equation}

\noindent we obtain

\begin{eqnarray}\label{Anomalous_Dimension}
&& (\gamma_\phi)_i{}^j(\alpha_0,\lambda_0) = \int \frac{d^4k}{(2\pi)^4}
\frac{1}{k^2} \frac{d}{dk^2} \Big( C(R)_i{}^j \frac{4
e^2}{R(k^2/\Lambda^2)} - \lambda^*_{imn} \lambda^{jmn}
\frac{4}{F^2(k^2/\Lambda^2)} \Big)\qquad\nonumber\\
&& + O(\alpha^2,\alpha\lambda^2,\lambda^4) = - C(R)_i{}^j
\frac{\alpha_0}{\pi} + \frac{1}{4\pi^2} \lambda^*_{0imn}
\lambda_0^{jmn} + O(\alpha_0^2,\alpha_0\lambda_0^2,\lambda_0^4).
\end{eqnarray}

\noindent Note that in the last equation the result is written in
terms of the bare coupling constants $\alpha_0$ and
$\lambda_0^{ijk}$, because we calculate the anomalous dimension
defined in terms of the bare charges. Certainly, the expression
(\ref{Anomalous_Dimension}) coincides with the well-known result.
However, it is interesting to compare the integral
(\ref{Anomalous_Dimension_Original}) with the integrals which give
the two-loop $\beta$-function with the considered regularization, as
it was done in, e.g., \cite{Kazantsev:2014yna}.

\subsection{One-loop renormalization of the quantum gauge superfield}
\hspace*{\parindent}\label{Subsection_One-Loop_V}

\begin{figure}[h]

\begin{picture}(0,4)
\put(2.6,2.4){\includegraphics[scale=0.4]{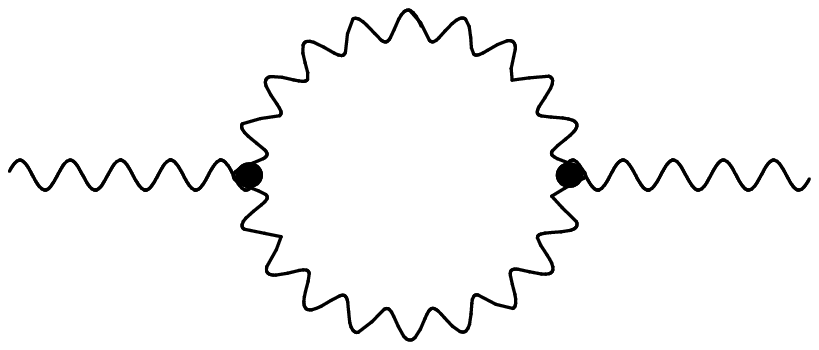}}
\put(3,0){\includegraphics[scale=0.4]{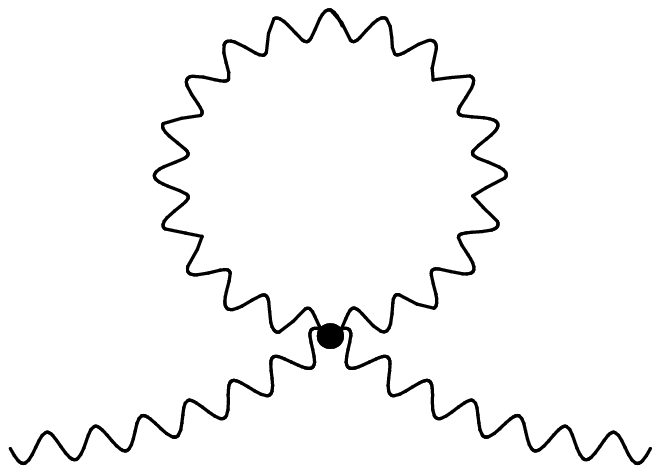}}
\put(6.4,2.43){\includegraphics[scale=0.4]{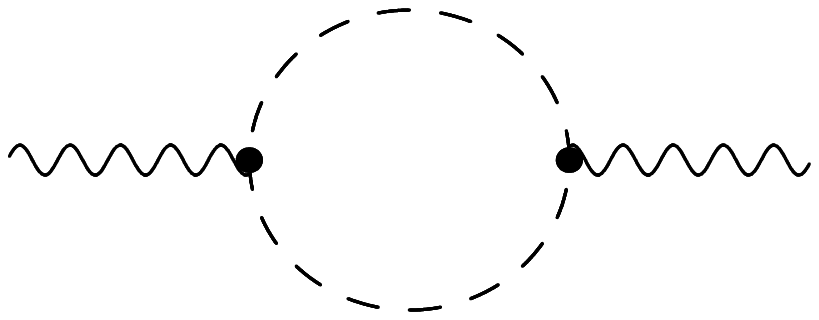}}
\put(6.7,-0.01){\includegraphics[scale=0.4]{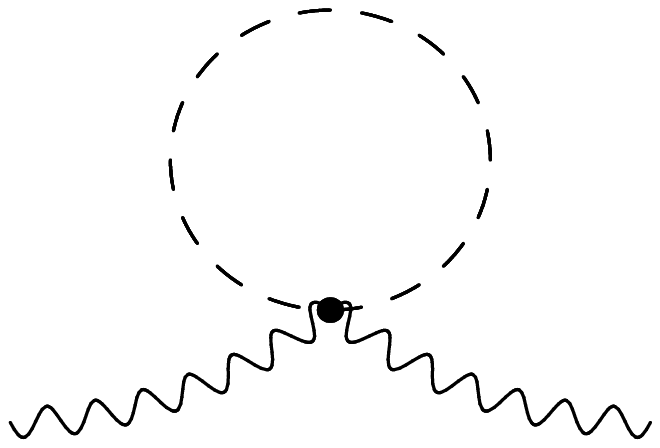}}
\put(10.2,2.4){\includegraphics[scale=0.4]{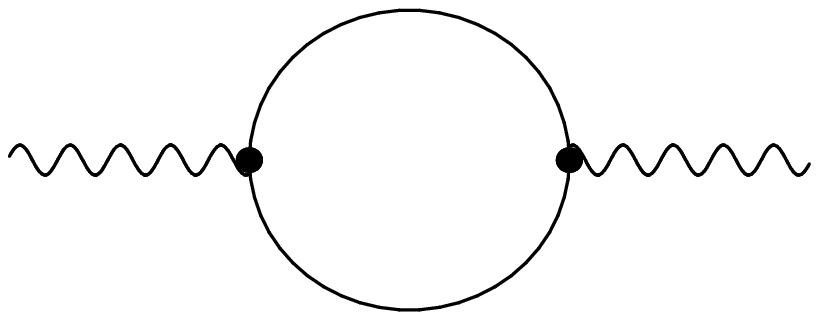}}
\put(10.6,-0.02){\includegraphics[scale=0.4]{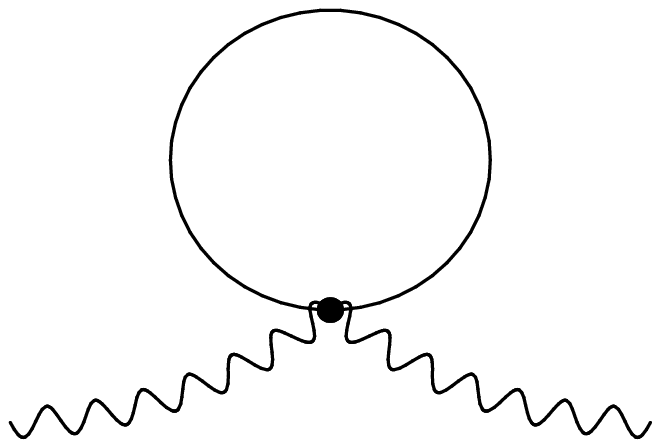}}
\end{picture}

\caption{One-loop diagrams contributing to the two-point Green
function of the quantum gauge superfield.}
\label{Figure_One_Loop_Quantum}
\end{figure}

From the two-point Green function of the quantum gauge superfield we
can find the constant $Z_V^2$. In the one-loop
approximation this Green function is contributed by the diagrams
presented in Fig. \ref{Figure_One_Loop_Quantum}. After calculating
them in the limit of the vanishing external momentum we obtained

\begin{eqnarray}
&& \frac{d\ln Z_V^2}{d\ln\Lambda} = \pi \alpha_0 \int
\frac{d^4q}{(2\pi)^4}
\frac{d}{d\ln\Lambda}\Bigg(\frac{\partial}{\partial q^\mu}
\frac{\partial}{\partial q_\mu} \Big[ \frac{C_2}{q^2}\ln
\Big(1+\frac{M_\varphi^2}{q^2 R^2}\Big) +\frac{2C_2}{q^2}
\ln\Big(1+\frac{M_\varphi^2}{q^2}\Big)\quad\nonumber\\
&& - \frac{T(R)}{q^2}\ln\Big(1+\frac{M_\Phi^2}{q^2 F^2}\Big)\Big] +
8 C_2\Big(- \frac{1}{3Rq^4} + \frac{\xi_0}{3K q^4}\Big) +
O(\alpha_0,\lambda_0^2) \Bigg).\qquad
\end{eqnarray}

\noindent Comparing this equation with the expression $d\ln
Z_\alpha/d\ln\Lambda$ given by Eq. (\ref{Gamma_Alpha}) we obtain

\begin{equation}\label{Gamma_V}
\gamma_V = - 4\pi \alpha_0 C_2\int\frac{d^4q}{(2\pi)^4}
\frac{d}{d\ln\Lambda} \Big(-\frac{1}{3R q^4} + \frac{\xi_0}{3 K
q^4}\Big)-\frac{1}{2}\cdot \frac{d\ln Z_\alpha}{d\ln\Lambda} + O(\alpha_0^2,\alpha_0\lambda_0^2).
\end{equation}

\noindent The integral in this expression can be easily calculated
by using Eq. (\ref{Derivative}):

\begin{eqnarray}
&& \int\frac{d^4q}{(2\pi)^4} \frac{d}{d\ln\Lambda} \Big(-\frac{1}{3R
q^4} + \frac{\xi_0}{3 K q^4}\Big)= - \int \frac{d^4q}{(2\pi)^4}
\frac{2}{3q^2} \frac{d}{dq^2}\Big(-\frac{1}{R} +
\frac{\xi_0}{K}\Big)\nonumber\\
&&\qquad\qquad\qquad\qquad\qquad\qquad\qquad\qquad\qquad =
-\frac{1}{24\pi^2} \Big(\frac{1}{R(0)} - \frac{\xi_0}{K(0)}\Big) =
-\frac{(1-\xi_0)}{24\pi^2},\qquad
\end{eqnarray}

\noindent so that finally we obtain

\begin{equation}\label{Gamma_V_Final}
\gamma_V = \frac{\alpha_0 C_2 (1-\xi_0)}{6\pi} + \frac{\beta(\alpha_0,\lambda_0)}{2\alpha_0}
+ O(\alpha_0^2,\alpha_0\lambda_0^2).
\end{equation}

\subsection{One-loop renormalization of the Faddeev--Popov ghosts}
\hspace*{\parindent}\label{Subsection_One-Loop_Ghosts}

\begin{figure}[h]
\begin{picture}(0,2)
\put(4.3,0.2){\includegraphics[scale=0.42]{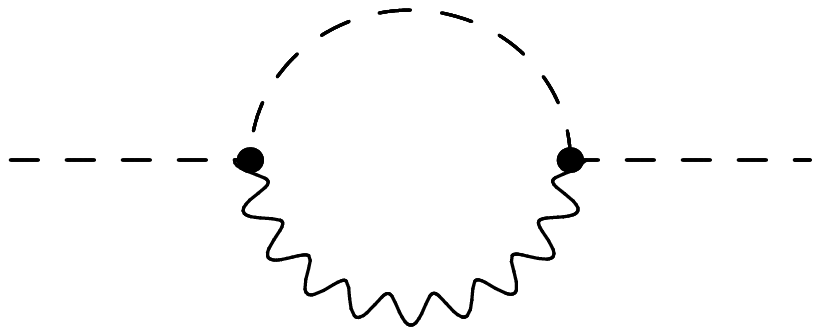}}
\put(8.7,0.0){\includegraphics[scale=0.41]{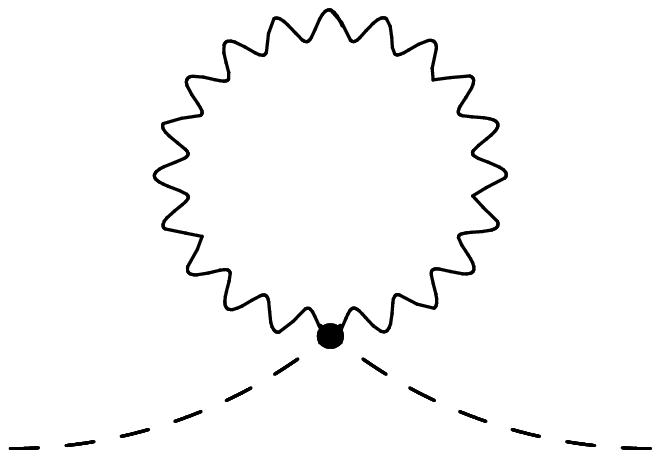}}
\end{picture}
\caption{Diagrams contributing to the one-loop anomalous dimension
of the Faddeev--Popov
ghosts.}\label{Figure_Ghost_Anomalous_Dimension}
\end{figure}

In order to find the anomalous dimension of the Faddeev--Popov
ghosts (defined in terms of the bare coupling constant) it is
necessary to calculate the diagrams presented in Fig.
\ref{Figure_Ghost_Anomalous_Dimension}. It is convenient to write
the result in the form

\begin{eqnarray}\label{Gamma_C}
&& \gamma_c = 4\pi
\alpha_0 C_2 \int\frac{d^4q}{(2\pi)^4} \frac{d}{d\ln\Lambda}
\Big(-\frac{1}{3R q^4} + \frac{\xi_0}{3 K q^4}\Big) +
O(\alpha_0^2,\alpha_0\lambda_0^2)
\nonumber\\
&&\qquad\qquad\qquad\qquad\qquad\qquad\qquad\qquad\qquad\quad = - \frac{\alpha_0 C_2
(1-\xi_0)}{6\pi} + O(\alpha_0^2,\alpha_0\lambda_0^2).\qquad
\end{eqnarray}

\noindent This implies that in the general gauge the two-point Green
function of the Faddeev--Popov ghosts is divergent, and the higher
covariant derivative regularization does regularize these
divergences. However, comparing Eq. (\ref{Gamma_C}) with Eq.
(\ref{Gamma_V}) (or Eq. (\ref{Gamma_V_Final})) we see that in the
one-loop approximation

\begin{equation}\label{Non_Renormalization}
\frac{d}{d\ln\Lambda}\Big(\ln Z_c + \ln Z_V -\frac{1}{2} \ln Z_\alpha\Big) = 0.
\end{equation}

\noindent As a consequence, the vertices of the type $\bar c\, V c$
are finite. Possibly, this statement is valid in all loops if the
regularization does not break the BRST invariance of the theory.
Note that for some particular regularizations and gauge fixing conditions
validity of Eq. (\ref{Non_Renormalization}) in the one-loop
approximation can also be seen from the results of Ref.
\cite{Slavnov:2003cx} for the pure ${\cal N}=1$ SYM theory and of
Ref. \cite{Buchbinder:2014wra,Buchbinder:2015eva} for the general ${\cal N}=2$ SYM
theory with matter.

\section{Conclusion}
\hspace*{\parindent}

In this paper we consider a general ${\cal N}=1$ SYM theory with
matter regularized by a very general version of the higher
derivative regularization which does not break the BRST invariance
and calculate all RG functions in the one-loop approximation. The
considered version of the higher derivative regularization was not
earlier used to obtain quantum corrections, because it leads to very
complicated calculations. However, it does not break symmetries of
the theory and seems to be very useful for the general derivation of
the NSVZ $\beta$-function in the non-Abelian case by the direct
summation of supergraphs. Making such a derivation one should consider
the one-loop approximation separately, and this problem is addressed
in this paper. In particular, we have demonstrated that
all one-loop momentum integrals for the $\beta$-function are
integrals of double total derivatives independently of the form of
the higher derivative term. This seems to be a general feature of
all supersymmetric theories.\footnote{The similar structures for
Abelian ${\cal N}=1$ supersymmetric theories regularized by the dimensional reduction
were considered in \cite{Aleshin:2015qqc}.} Certainly, the result of the
calculation coincided with the well-known expression for the
one-loop $\beta$-function in the supersymmetric case. Also we have
obtained the momentum integrals defining the one-loop anomalous
dimension, which also coincided with the well-known expression. In
prospect, these integrals can be compared with integrals giving the
two-loop $\beta$-function, which are related to them due to the
existence of the NSVZ $\beta$-function. Also we obtained that the
vertices $\bar c\, V c$, $\bar c\, V c^+$, $\bar c^+ V c$, and $\bar
c^+ V c^+$ (containing two lines of the Faddeev--Popov ghosts and a
single line of the quantum gauge superfield) are not renormalized in
the considered approximation. Possibly, this feature is valid in an
arbitrary order of the perturbation theory.

\bigskip

\section*{Acknowledgments}
\hspace*{\parindent}

The authors are very grateful to D.S.Kolupaev for valuable
discussions. The work of K.S. is supported by the RFBR grant No.
14-01-00695.


\end{document}